    \documentclass[fleqn,usenatbib]{mnras}
    \usepackage{amssymb}
    \usepackage{graphicx}
    \usepackage{orcidlink}
    
    \usepackage[table]{xcolor}
    \usepackage{amsmath}

    \newcommand{\msun}{\mathrm{M_\odot}}
    \newcommand{\mach}{\mathcal{M}}
    \newcommand{\avir}{\alpha_\mathrm{vir}}
    \newcommand{\au}{\mathrm{AU}}
    \newcommand{\fdl}{f_{\Delta L}}
    \newcommand{\lnplr}{\ln(\mathrm{PLR})}
    \newcommand{\myr}{\mathrm{Myr}}

    \definecolor{mygreen}{RGB}{34,139,34}

       \title[Stellar spin in very young clusters]{Stellar spin in young star clusters: comparison between  simulations and observations}
    
       \author[G.~Cordoni et al.]{Giacomo Cordoni\,\orcidlink{0000-0002-7690-7683}$^{1}$\thanks{E-mail: giacomo.cordoni@anu.edu.au},
              Sajay Sunny Mathew\,\orcidlink{0000-0002-8381-8195}$^{1}$, and
              Christoph Federrath\,\orcidlink{0000-0002-0706-2306}$^{1}$
                  \\
        $^{1}$Research School of Astronomy and Astrophysics, Australian National University, Canberra, ACT~2611, Australia
        }
    
    \date{Accepted XXX. Received YYY; in original form ZZZ}
    
    \pubyear{2025}
    
\begin{document}
    \label{firstpage}
    \maketitle
    
    \begin{abstract}

    The angular momentum evolution of stars is crucial for understanding the formation and evolution of stars and star clusters. Using high-resolution magnetohydrodynamical (MHD) simulations of star formation in clouds with different physical properties, we study the initial distribution of stellar rotation periods in young clusters. We compare these results with observations of young Galactic clusters. Simulations qualitatively reproduce the observed trend of increasing rotation period with stellar mass. Additionally, simulations with lower virial parameter (ratio of turbulence to gravity) or solenoidal turbulence driving produce period-mass distributions that more closely match the observed ones. These simulations also recover the break in the mass-period relation. However, the break appears at higher masses than in observations and is absent in the youngest simulated clusters. This suggests that the emergence of the break is an important diagnostic of angular momentum evolution during the earliest stages of cluster formation.
    
    The simulations yield stars that rotate about an order of magnitude faster than those observed. This discrepancy mainly reflects the earlier evolutionary stage of the simulations, while unresolved physical interactions between stars and discs might also contribute. This conclusion is supported by simulations showing a significant period increase within $0.1-1\,\myr$. We quantify the required angular-momentum loss by rescaling simulated rotation periods to match observations, finding that $80-95\%$ of the initial angular momentum must be removed within the first $\myr$. Our results highlight that understanding the earliest stages of star cluster formation is fundamental to addressing the angular momentum problem.

    \end{abstract}
    
    %
    
    \section{Introduction} \label{sec:intro}
    The angular momentum content and rotation of stars are fundamental properties that influence their structure, evolution, and interaction with their environment \citep[e.g.,][]{maeder2009}. Rotation affects stellar internal structure, mass-loss rates, magnetic field generation, and high-energy emission \citep[e.g.,][]{wright2011, brun2017}, with important implications for the evolution of planetary systems. For low-mass stars, the rotational evolution begins in the earliest phases of star formation, when the protostar is still embedded in an envelope and accreting through a circumstellar disc. At this stage, a combination of processes—including magnetic star–disc coupling, stellar winds and jets, and magnetospheric ejections—can remove a large fraction of the initial angular momentum \citep[e.g.,][]{matt2005,Banerjee2006,zanni2013}.

    Observations of very young clusters (approximately younger than $3\,\myr$) already reveal a wide spread of stellar rotation periods, ranging from less than a day to more than ten days \citep[e.g.][]{rebull2014, venuti2017, serna2021}. This early spread sets the initial conditions for later spin evolution, yet its physical origin remains poorly constrained. In particular, most stellar spin evolution models and simulations adopt the observed distribution at $\sim 1\,\myr$ as an input, without directly addressing how it emerges from the star formation process \citep[e.g.][]{bouvier2014, gallet2019}. As a result, the link between the initial angular momentum content of molecular cloud cores and the distribution of stellar spins at the start of the pre-main-sequence remains an open question. Nevertheless, theory and observations indicate that substantial angular-momentum loss must already occur very early, from core collapse through the Class~0/I/II stages via magnetic braking, jets/outflows, winds, and star–disc coupling (see e.g.~the review by \citealt{bodenheimer1995}).

    State-of-the-art magnetohydrodynamical (MHD) simulations now include a wide range of relevant physics, including gravity, turbulence, magnetic fields, radiative feedback, accretion, and protostellar evolution \citep[e.g.][]{Federrath2015,mathew2021,AppelEtAl2023}—and have begun to produce synthetic distributions of stellar angular momentum \citep[see e.g.,][]{mathew2023}. However, these predicted distributions have rarely been compared directly with observations of very young clusters, and the degree to which simulations can reproduce the observed spread remains unclear. Additionally, as simulations of star clusters formation typically follow the first $0.1\,\myr$ of star formation, they can provide a direct estimate of the initial period distributions, as determined directly by star formation processes.

    In this work, we build upon the simulations of \citet{mathew2021, mathew2025} and explore the stellar rotation period distributions predicted by the simulations under different physical conditions. Our goal is to bridge the gap by comparing the rotation period distributions produced by these MHD simulations to those measured in very young clusters. Specifically, we aim to assess whether the initial spin distribution emerging from the simulations is consistent with observations, and to identify which physical mechanisms, such as turbulence, accretion history, magnetic coupling, shape the early evolution of stellar angular momentum. Additionally, the comparison will provide realistic estimates of the initial spin distribution.

    The paper is organised as follows: in Sec.~\ref{sec:sim} we introduce the MHD simulations and their properties, briefly describing the numerical methods and the main key physical parameters. In Sec.~\ref{sec:obs} we describe the observational datasets included in the comparison and how the observational periods have been derived. Finally, the qualitative and quantitative comparison between simulations and observations is discussed in Sec.~\ref{sec:discussion}, while Sec.~\ref{sec:conclusion} presents the summary and conclusion.
    
    \section{Simulations} \label{sec:sim}
    
    \subsection{Simulation methods and parameters}
    
    \begin{table*}
        \centering
        \renewcommand{\arraystretch}{1.2}
        \setlength{\tabcolsep}{7.0pt}
        \begin{tabular}{cccccccccc}
            \hline
            & $\alpha_{\rm vir}$ & $\mathcal{M}$ & $\zeta$ & $\bar{M}_\star$ & Range $M_\star$ & $16^\mathrm{th}\!-\!84^\mathrm{th}\,M_\star$ & $\bar{P}_\mathrm{rot}$ & Range $P_\mathrm{rot}$ & $16^\mathrm{th}\!-\!84^\mathrm{th}\,P_\mathrm{rot}$ \\
            & & & & [$\msun$] & [$\msun$] & [$\msun$] & [days] & [days] & [days] \\
            \hline
      \textbf{1} & $0.50$ & $2.50$  & $0.50$ & $0.71$ & $0.01$--$6.01$  & $0.04$--$1.88$  & $0.22$ & $0.00$--$3.58$   & $0.04$--$0.87$ \\
      \textbf{2} & $0.50$ & $10.00$ & $0.50$ & $0.26$ & $0.02$--$4.15$  & $0.08$--$0.80$  & $0.27$ & $0.02$--$29.13$  & $0.11$--$1.01$ \\
      \textbf{3} & $0.50$ & $5.00$  & $0.00$ & $0.24$ & $0.01$--$3.88$  & $0.07$--$0.70$  & $0.41$ & $0.01$--$41.76$  & $0.15$--$1.30$ \\
      \textbf{4} & $0.50$ & $5.00$  & $0.50$ & $0.38$ & $0.01$--$4.83$  & $0.09$--$1.09$  & $0.25$ & $0.01$--$51.24$  & $0.09$--$0.89$ \\
      \textbf{5} & $0.50$ & $5.00$  & $1.00$ & $0.40$ & $0.01$--$5.47$  & $0.12$--$1.25$  & $0.23$ & $0.00$--$4.49$   & $0.09$--$0.85$ \\
      \textbf{6} & $0.06$ & $5.00$  & $0.50$ & $0.19$ & $0.01$--$3.46$  & $0.04$--$0.79$  & $0.32$ & $0.02$--$172.42$ & $0.13$--$1.11$ \\
      \textbf{7} & $0.13$ & $5.00$  & $0.50$ & $0.23$ & $0.01$--$3.85$  & $0.07$--$0.76$  & $0.29$ & $0.01$--$230.73$ & $0.12$--$1.27$ \\
            \hline
        \end{tabular}
        \caption{Simulation parameters and results -- in order of columns: virial parameter $\avir$, sonic Mach number $\mach$, turbulence driving parameter $\zeta$, average stellar mass, stellar mass full range and percentile range, average stellar rotation period, and stellar period full range and percentile range.}
        \label{tab:tab sim}
    \end{table*}

    Here we provide only a brief and general description of the simulations used in this work, and we refer to \citet{mathew2021,mathew2025} for a detailed description of the numerical recipes adopted. In a nutshell, cluster formation is modelled by solving MHD equations in the presence of gravity with a modified version of the FLASH (version~4) code \citep{fryxell2000,DubeyEtAl2008} using the HLL5R Riemann solver by \citet{waagan2011}. The simulations implement turbulence, protostellar heating and jets/outflows, in addition to gravity and magnetic fields. 
    
    Star formation in the simulations is treated via the sink-particle technique introduced by \citet{federrath2010b}. Initially, the computational domain contains only gas; no sinks or protostars are imposed. As the turbulent cloud evolves, local regions may become gravitationally bound and start collapsing. After passing a series of checks for local collapse, a sink particle is created. The excess gas mass above the sink particle density threshold, dynamically set by the Jeans resolution criterion, is removed from the gas and converted into sink mass. This procedure ensures that sinks form self-consistently and only in collapsing regions, avoiding artificial fragmentation \citep[see detailed criteria and tests in][]{federrath2010b}. Once created, a sink particle can accrete gas from its surroundings whenever neighbouring cells satisfy the same collapse conditions and are gravitationally bound to the sink particle. The sink particle inherits the accreted mass, and linear and angular momentum, such that each sink represents an unresolved star-plus-disc system whose mass, momentum, and spin evolve through continued accretion. Protostellar feedback is coupled to each sink particle through the sub-grid jet/outflow and radiative-heating prescriptions described in \citet{federrath2014} and \citet{mathew2020}, respectively.
    
    The finest grid spacing is $100\,\au$, and therefore, while fragmentation on extended-disc scales can occur, fragmentation on typical disc scales is not fully resolved. Protostellar jet and outflow feedback \citep[following the implementation by][]{federrath2014} and heating \citep[see implementation details in][]{mathew2020} are included via sub-resolution models. We note that $90\%$ of the angular momentum accreted onto the sink particles is redistributed to the jet/outflow components, based on the observations in \citet{Bacciotti2002} and numerical works, e.g., \citet{Banerjee2006} and \citet{Hennebelle2008}. 

    The simulations are evolved until the star formation efficiency reaches $\approx 5\%$, that is, when $5\%$ of the initial cloud mass has been converted into stars, at which point the oldest stellar ages are typically $\sim0.5-1\,\myr$. 

    Since our goal is to compare the modelled clusters, specifically their stellar rotation, to observations, we explore families of simulations with different physical conditions by varying the virial parameter $(\avir)$, the sonic Mach number $(\mach)$, and the turbulence driving mode $(\zeta)$. The cloud parameters adopted for each simulation are listed in Tab.~\ref{tab:tab sim}. We provide a brief outline of parameters and their expected impact:
    
    \begin{itemize}
    
    \item The \textbf{virial parameter} $\avir=2E_\mathrm{turb}/|E_\mathrm{grav}|$ \citep{BertoldiMcKee1992,FederrathKlessen2012}, controls the ratio of turbulent kinetic to gravitational energy. Systems with low $\avir$ are more gravitationally bound and form stars more efficiently and at lower characteristic masses, while higher values delay or suppress collapse, shifting the mass scale \citep{Haugb_lle_2018,2018A&A...611A..88L,mathew2025}. Here we explore three values of $\avir=(0.0625,\,0.125,\,0.5)$ where the latter represents a typical value, although observations suggest a wide distribution of $\avir$ \citep[e.g.,][]{KauffmannPillaiGoldsmith2013}.
    
    \item The sonic \textbf{Mach number} $\mach$ is the ratio of turbulent velocity dispersion to sound speed. Higher $\mach$ results in stronger compressions, producing shocks when $\mach>1$. Typical molecular cloud values are in the range $\mach=1$--$20$ \citep[e.g.,][]{SchneiderEtAl2013,FederrathEtAl2016}

    \item The \textbf{turbulence driving mode} parameter $\zeta$, implemented as a stochastic forcing \citep[an Ornstein–Uhlenbeck process,][]{eswaran1988}, represents the ratio of energy in solenoidal (divergence-free) to compressive (curl-free) modes. Variation in $\zeta$ is found to directly alter collapse statistics, fragmentation, binary properties, and angular-momentum acquisition by sinks \citep{federrath2008,federrath2010a,FederrathKlessen2012,mathew2023, mathew2024}. We compare simulations with $\zeta=0$, $0.5$, and $1$, corresponding to compressive, mixed, and solenoidal driving, respectively.

    \end{itemize}
    
    In this work we have also applied a post-processing correction to each raw sink mass in order to remove the unresolved disc contribution in the angular momentum and period calculations. Indeed, as discussed in \citet{mathew2021, mathew2025}, the sink radius includes both star and disc, so that the total mass is not only stellar mass. 

    Observational surveys of young stellar clusters and associations show that the fraction of stars hosting optically thick inner discs falls below half by cluster ages of about $3\,\myr$, and is essentially zero by around $10\,\myr$ \citep[e.g.][]{mamajek2009, ribas2015}.
    Additionally, higher-mass stars tend to disperse their discs substantially faster than lower-mass ones \citep{ribas2015}. We therefore subtract from each sink a disc mass that exponentially decays with stellar age, and that also depends on stellar mass. The correction removes up to 50\% of the initial system mass for the youngest objects and approximately 10\% for the oldest, most massive stars. We include in Appendix~\ref{appendix:disc} a detail description of the adopted mass-loss model, as well as tests of different model details. 

    Finally, for all stars in the simulations, we derive their rotational periods from the total angular momentum, assuming the moment of inertia $I$ of a uniform sphere. Thus, we take the magnitude of the angular–momentum vector $L$, adopted $I = k^2 M_\star R_\star^2$ with $k^2=0.4$ for a uniform sphere, and compute the spin period $P =2\pi I/L$. While this is a crude approximation to stellar structure, we also tested alternatives and find our conclusions unchanged. Specifically, we re-compute the rotation periods using $k^2=0.2$, similar to the values found in \citet{claret1989} for stars in our age range.

    \subsection{Results and discussion}

    \begin{figure*}
        \centering
        \includegraphics[width=0.99\textwidth]{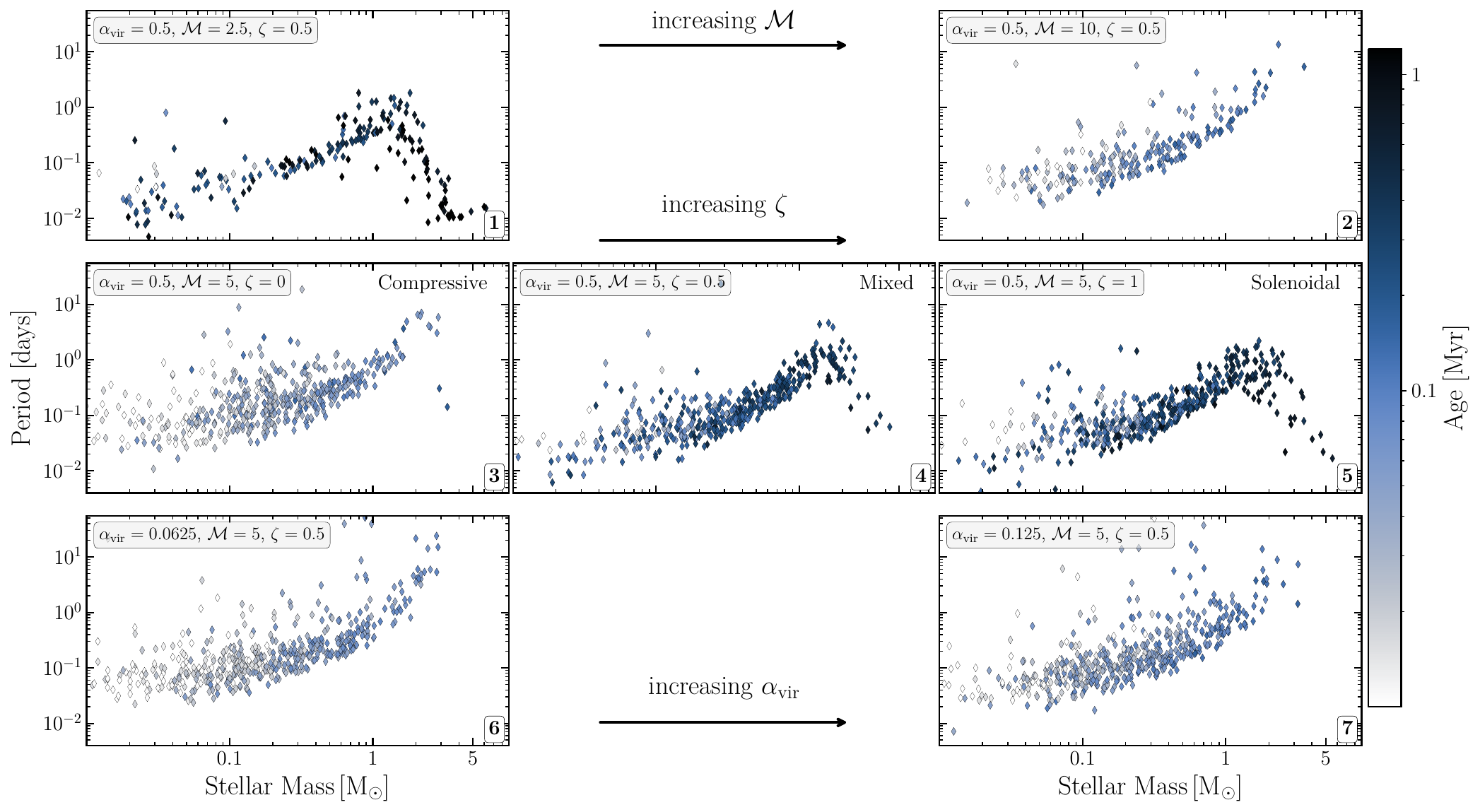}
        \caption{Stellar rotation period vs.~stellar mass for the numerical simulations listed in Tab.~\ref{tab:tab sim} (panel numbers correspond to simulation numbers). Stars are coloured according to their age, as defined in Sec.~\ref{sec:sim} and indicated in the colour map. All simulations show the overall trend of longer periods at higher masses; several setups (e.g.~1, 4, and 5) also exhibit a high-mass break after which stars start to rotate faster. This break is interpreted as runaway accretion driving spin-up of massive stars. Simulations with younger stars, e.g.~lighter colours, instead do not show evidence of the break. The break in stellar rotation vs. mass is also observed in young star clusters, albeit at lower stellar masses.}
        \label{fig:sim}
    \end{figure*}

    Figure~\ref{fig:sim} shows the stellar rotation period as a function of mass for our seven simulation models (see Tab.~\ref{tab:tab sim}).
    
    The simulation parameter set is indicated in the top-left corner of each panel, where we specify $\avir$, $\mach$, and $\zeta$. The ``base'' model $(\avir=0.5,\,\mach=5,\,\zeta=0.5)$ sits in the central panel. Points are coloured by stellar age (defined as the time between star formation and the end of the simulation), with white denoting the youngest objects and black the oldest, as indicated in the colour map.
    
    All simulations show the same overall trend of increasing rotation period with stellar mass, with a narrower spread at the low-mass end. In simulations~1, 4, and~5, a clear break appears at high masses, after which the periods decrease. This feature is consistent with runaway accretion onto massive stars, which causes spin-up (see Sec.~\ref{sec:discussion} for details). These simulations also include slightly older stars, between 0.5 and $1\,\myr$, and stars with lower rotation periods. By contrast, no break is seen in the simulations that produce younger clusters, where the angular momentum evolution has not yet progressed as far.

    The simulations also yield specific angular momenta consistent with those obtained in earlier studies. In particular, \citet{mathew2021} compared the angular momentum distribution from this simulation suite with ALMA measurements of the Class~0 protostar B335 \citep{yen2015}, finding good agreement on scales between of a few hundreds $\au$. This consistency supports the reliability of the angular-momentum treatment in the simulations at protostellar scales.
    
    Relevant properties, such as stellar mass and period statistics are listed in Tab.~\ref{tab:tab sim}, where the first column is the simulation model, matched in order with the panel number in Fig.~\ref{fig:sim}.

    \section{Observations} \label{sec:obs}
    
    Over the past decades, major advances in observational astronomy have transformed our understanding of stellar rotation. High-precision, continuous light curves from space missions such as Kepler \citep[][]{borucki2010kepler}, K2 \citep[][]{howell2014k2}, and the Transiting Exoplanet Survey Satellite (TESS; \citealt{ricker2015tess}), together with wide-field monitoring from ground-based facilities like the Zwicky Transient Facility \citep[ZTF;][]{bellm2019ztf}, have provided unprecedented data quality. These datasets have enabled the measurement of rotation periods for tens of thousands of stars, establishing stellar spin as a central parameter for stellar physics and its time evolution \citep[e.g.,][]{rampalli2021, reinhold2020, douglas2019, lu2022}.
    
    Rotation periods are typically obtained from quasi-periodic brightness variations caused by starspots modulated in and out of view. Observations across clusters of different ages reveal systematic trends: stars spin up during their pre-main-sequence contraction and spin down during the main sequence as angular momentum is shed through magnetised winds \citep[e.g.,][]{venuti2017, rebull2018, rebull2020, rebull2022, healy2020, healy2021, healy2023, godoyrivera2021}. At the youngest ages ($<5\,\myr$), stars already show a wide dispersion in rotation rates, the origin of which remains debated \citep{gallet2019}. Young open clusters are therefore crucial benchmarks: they provide coeval stellar populations with well-defined ages and distances, enabling direct comparison to models of angular momentum evolution.
    
    \subsection{Literature data}
    
    In this work we compile published rotation period samples from several key surveys and cluster studies, and compare them against our MHD simulations. Stellar masses were taken from \citet{hunt2024} whenever available. When not, we derived masses by interpolating PARSEC isochrones \citep{marigo2017} at the cluster ages and distances listed in Table~\ref{tab:tab cl}. For clusters in \citet{rebull2018, rebull2020, rebull2022}, we used the distances listed in Table~\ref{tab:tab cl}, or the individual distances provided in the respective datasets when available (e.g.~in \citealt{rebull2022}). For \citet{venuti2017} and \citet{getman2023}, stellar masses were already provided, and we refer directly to their methodology. Further details on the used datasets are summarised below.
    
    \subsubsection{Healy et al.~(2023)}
    
    \citet{healy2023} analysed rotation and inclination distributions in 11~open clusters spanning $5-500\,\myr$. Rotation periods were measured from TESS light curves processed with the \texttt{PATHOS} pipeline \citep{nardiello2019} and the \texttt{eleanor} software \citep{feinstein2019}, using both autocorrelation functions and Lomb–Scargle periodograms. Stellar masses for this work were taken from \citet{hunt2024}, who derived them by interpolating PARSEC isochrones at homogeneous cluster parameters \citep{hunt2024}. We verified that their reported cluster parameters produce consistent masses. Finally, we only include clusters younger than $200\,\myr$, thus excluding NGC\,2548. 
    
    \subsubsection{Getman et al.~(2023)}
    
    \citet{getman2023} studied rotation in six clusters between 7 and $25\,\myr$: NGC\,2362 ($5\,\myr$),NGC\,1502 and NGC\,2169 ($7\,\myr$), NGC\,869 and NGC\,884 ($13\,\myr$; $h$ and $\chi$\,Persei), NGC\,1960 ($22\,\myr$), and NGC\,2232 ($25\,\myr$). Candidate members were identified using a combination of Chandra X-ray detections and Gaia astrometry, biasing the sample towards magnetically active stars. Rotation periods were derived from ZTF DR10 $g,r$ light curves using Lomb–Scargle and autocorrelation analyses, with injected-signal simulations employed to remove aliases. This yielded 471 reliable periods. Stellar masses were provided by the authors, and we refer to \citet{getman2023} for details of their derivation. The main caveats remain the activity-based selection (which disfavours slow rotators) and the daily aliasing of ground-based monitoring near $P\sim1$~d.

    \subsubsection{Venuti et al.~2017} 

    \citet{venuti2017} characterised the rotation properties of young stars in the star-forming region NGC\,2264 ($\sim3\,\myr$), with the goal of investigating the connection between accretion and stellar angular momentum at an age when roughly half of the stars have already lost their discs. Their analysis combined optical photometric monitoring with \emph{CoRoT} over a 38~day baseline, yielding high-cadence light curves for about 500~cluster members. Rotation periods were determined with three independent methods: the Lomb–Scargle periodogram, the autocorrelation function, and the string-length algorithm \citep{dworetsky1983}. Period detections were validated through $Q$-statistics \citep{cody2014}, and visual inspection of both direct and phase-folded light curves. Reliable periods were obtained for $\sim 300$ stars of the monitored sample, which are included in this work \citep[Table~4 of][]{venuti2017}.

    \subsubsection{Rebull et al.~(2018, 2020, 2022)}
    
    For the youngest clusters, we include $\rho$\,Oph ($1\,\myr$) and Upper Scorpius ($8\,\myr$) from \citet{rebull2018}, Taurus ($3\,\myr$) from \citet{rebull2020}, and Upper Centaurus–Lupus ($16\,\myr$) and Lower Centaurus–Crux ($17\,\myr$) from \citet{rebull2022}. Periods in \citet{rebull2018, rebull2020} were measured from \emph{K2} light curves, while \citet{rebull2022} analysed TESS data. Periods were identified using Lomb–Scargle periodograms and vetted visually to remove blended or non-rotational variability (eclipses, pulsations). For TESS targets, the authors relied on the best available light curves, including extractions from \texttt{eleanor} \citep{feinstein2019}, the Cluster Difference Imaging Photometric Survey \citep[CDIPS;][]{bouma2019}, and the MIT Quick-Look Pipeline (QLP; \citealt{huang2020a, huang2020b}). Membership reliability was categorised as ``gold'', ``silver'', or ``bronze''; here we restrict to the ``gold'' members.
    
    Stellar masses for these clusters were derived by interpolating PARSEC isochrones at the reported ages and distances, as listed in Table~\ref{tab:tab cl}. When available, we adopted the individual stellar distances indicated by the authors (e.g.~\citealt{rebull2022}, based on Gaia DR2 parallaxes from \citealt{bailerjones2018})\footnote{See Sec.~3.5 of \citet{rebull2022} for a detailed justification of this choice.}. Mass estimates rely primarily on 2MASS $V-K_\mathrm{s}$ colours, which can introduce uncertainties in the pre-main-sequence regime due to wide colour spreads; to mitigate this, masses were derived separately from $V$ and $K_\mathrm{s}$ magnitudes and averaged.
    
    It is worth noting a few caveats regarding these datasets. At very young ages, disc-related and accretion-driven variability reduces the measurable period fraction. The limited baselines of \emph{K2} ($\sim$70–80~d) and TESS (27~d per sector) restrict sensitivity to long periods ($\gtrsim30$~d). TESS's large pixel scale increases blending risks, partially addressed by the membership grading but not entirely avoidable. In addition, many of the stars in $\rho$\,Oph, Upper Scorpius, Taurus, and UCL/LCC are still in the pre-main-sequence phase, which naturally contributes to the broad dispersion in measured periods and to the challenges of deriving accurate stellar parameters. Notably, in UCL/LCC, disc-bearing M~dwarfs show a concentration near $P\sim2$~d, likely reflecting ongoing disc–rotation interactions.

    \begin{table*}
        \centering
        \renewcommand{\arraystretch}{1.2} 
        \setlength{\tabcolsep}{7.0pt} 
        \begin{tabular}{ccccccccccc}
        \hline
         & Cluster     & Age & Dist. & Source 
         & $\bar{M}_\star$      & Range $M_\star$        & $16^\mathrm{th}\!-\!84^\mathrm{th}\,M_\star$ 
         & $\bar{P}_\mathrm{rot}$ & Range $P_\mathrm{rot}$  & $16^\mathrm{th}\!-\!84^\mathrm{th}\,P_\mathrm{rot}$ \\
         &                 & [$\myr$]             & [pc]                   & 
         & [$\msun$]        & [$\msun$]         & [$\msun$] 
         & [days]             & [days]               & [days] \\
         \hline
      \textbf{1}  & $\rho$\,Oph        & $1.0$    & 139 & $1$  & $1.00$ & $0.14$--$7.2$  & $0.44$--$1.80$  & $3.21$ & $0.28$--$14.51$ & $1.24$--$6.40$ \\
      \textbf{2}  & NGC\,2264          & $3.0$    & 760 & $6$  & $0.69$ & $0.13$--$2.72$   & $0.31$--$1.40$  & $3.82$ & $0.08$--$19.50$ & $1.22$--$8.64$ \\
      \textbf{3}  & Taurus             & $3.0$    & 138 & $2$  & $0.74$ & $0.09$--$5.20$   & $0.31$--$1.56$  & $2.99$ & $0.47$--$27.22$ & $1.51$--$6.68$ \\
      \textbf{4}  & Collinder\,69      & $4.9$    & 394 & $3$  & $0.55$ & $0.22$--$5.58$   & $0.33$--$0.94$  & $2.91$ & $0.43$--$14.60$ & $1.25$--$6.02$ \\
      \textbf{5}  & NGC\,2362          & $5.0$    & 1228 & $4$  & $0.64$ & $0.27$--$1.15$   & $0.45$--$0.79$  & $4.53$ & $0.32$--$27.01$ & $1.24$--$8.10$ \\
      \textbf{6}  & NGC\,1502          & $7.0$    & 1024 & $4$  & $0.89$ & $0.51$--$3.55$   & $0.69$--$1.27$  & $3.70$ & $0.29$--$22.44$ & $0.89$--$10.30$ \\
      \textbf{7} & NGC\,2169          & $7.0$    & 935 & $4$  & $0.76$ & $0.52$--$1.69$   & $0.62$--$1.10$  & $4.59$ & $0.72$--$20.22$ & $1.20$--$9.61$ \\
      \textbf{8}  & ASCC\,16           & $7.1$    & 342 & $3$  & $0.45$ & $0.22$--$1.30$   & $0.35$--$0.91$  & $2.73$ & $0.46$--$11.10$ & $1.15$--$5.14$ \\
      \textbf{9}  & ASCC\,19           & $8.0$    & 352 & $3$  & $0.64$ & $0.40$--$1.47$   & $0.54$--$0.89$  & $2.88$ & $0.34$--$12.40$ & $1.36$--$6.04$ \\
      \textbf{10}  & Upper Scorpius     & $8.0$    & 145 & $1$  & $0.52$ & $0.09$--$5.90$   & $0.21$--$1.25$  & $1.94$ & $0.05$--$38.39$ & $0.76$--$5.73$ \\
      \textbf{11}  & Gulliver\,6        & $8.6$    & 408 & $3$  & $0.54$ & $0.24$--$1.62$   & $0.34$--$1.00$  & $3.49$ & $0.36$--$11.30$ & $1.08$--$5.50$ \\
      \textbf{12} & Pozzo\,1           & $10.2$   & 347 & $3$  & $0.70$ & $0.27$--$1.54$   & $0.39$--$1.26$  & $2.78$ & $0.46$--$9.00$  & $1.09$--$5.83$ \\
      \textbf{13} & BH\,56             & $11.8$   & 880 & $3$  & $1.27$ & $0.64$--$2.36$   & $0.72$--$1.53$  & $2.87$ & $0.38$--$7.51$  & $0.99$--$5.02$ \\
      \textbf{14} & NGC\,869           & $13.0$   & 2280 & $4$  & $1.18$ & $0.74$--$6.45$   & $1.01$--$1.33$  & $0.89$ & $0.22$--$9.56$  & $0.42$--$4.01$ \\
      \textbf{15} & NGC\,884           & $13.0$   & 2293 & $4$  & $1.18$ & $0.73$--$5.80$   & $1.01$--$1.34$  & $0.88$ & $0.24$--$11.23$ & $0.40$--$3.91$ \\
      \textbf{16} & UCL/LCC            & $16.0$   & 130 & $5$  & $0.40$ & $0.09$--$3.13$   & $0.21$--$0.82$  & $1.67$ & $0.06$--$17.06$ & $0.59$--$4.73$ \\
      \textbf{17} & NGC\,1960          & $22.0$   & 1125 & $4$  & $0.83$ & $0.60$--$1.29$   & $0.70$--$1.00$  & $1.92$ & $0.26$--$8.90$  & $0.72$--$5.73$ \\
      \textbf{18} & NGC\,2232          & $25.0$   & 318 & $4$  & $0.60$ & $0.37$--$1.06$   & $0.42$--$0.91$  & $1.53$ & $0.28$--$32.51$ & $0.53$--$3.08$ \\
      \textbf{19} & NGC\,2547          & $31.7$   & 382 & $3$  & $0.87$ & $0.34$--$1.97$   & $0.54$--$1.28$  & $2.10$ & $0.24$--$6.65$  & $0.78$--$4.89$ \\
      \textbf{20} & Alpha\,Persei      & $58.3$   & 174 & $3$  & $0.66$ & $0.20$--$3.76$   & $0.35$--$1.12$  & $1.39$ & $0.22$--$7.80$  & $0.43$--$4.00$ \\
      \textbf{21} & Blanco\,1          & $105.0$  & 234 & $3$  & $0.57$ & $0.29$--$2.58$   & $0.37$--$1.03$  & $1.80$ & $0.26$--$10.34$ & $0.73$--$5.74$ \\
      \textbf{22} & NGC\,2516          & $140.0$  & 407 & $3$  & $0.97$ & $0.48$--$3.25$   & $0.61$--$1.19$  & $2.45$ & $0.22$--$7.30$  & $0.69$--$4.81$ \\
      \textbf{23} & NGC\,2422          & $155.9$  & 468 & $3$  & $0.85$ & $0.46$--$4.53$   & $0.60$--$1.21$  & $3.07$ & $0.27$--$9.50$  & $0.82$--$6.40$ \\
            \hline
        \end{tabular}
        \caption{ Cluster properties: cluster name, age, distance, source: (1)-\citet{rebull2018}, (2)-\citet{rebull2020}, (3)-\citet{healy2023}, (4)-\citet{getman2023}, (5)-\citet{rebull2022}, (6)-\citet{venuti2017}, average stellar mass, range of stellar masses and 16$^\mathrm{th}$--84$^\mathrm{th}$~percentiles, and same for rotation periods.}
        \label{tab:tab cl}
    \end{table*}

    \subsection{Results and discussion}

\begin{figure*}
        \centering
        \includegraphics[width=0.9\textwidth]{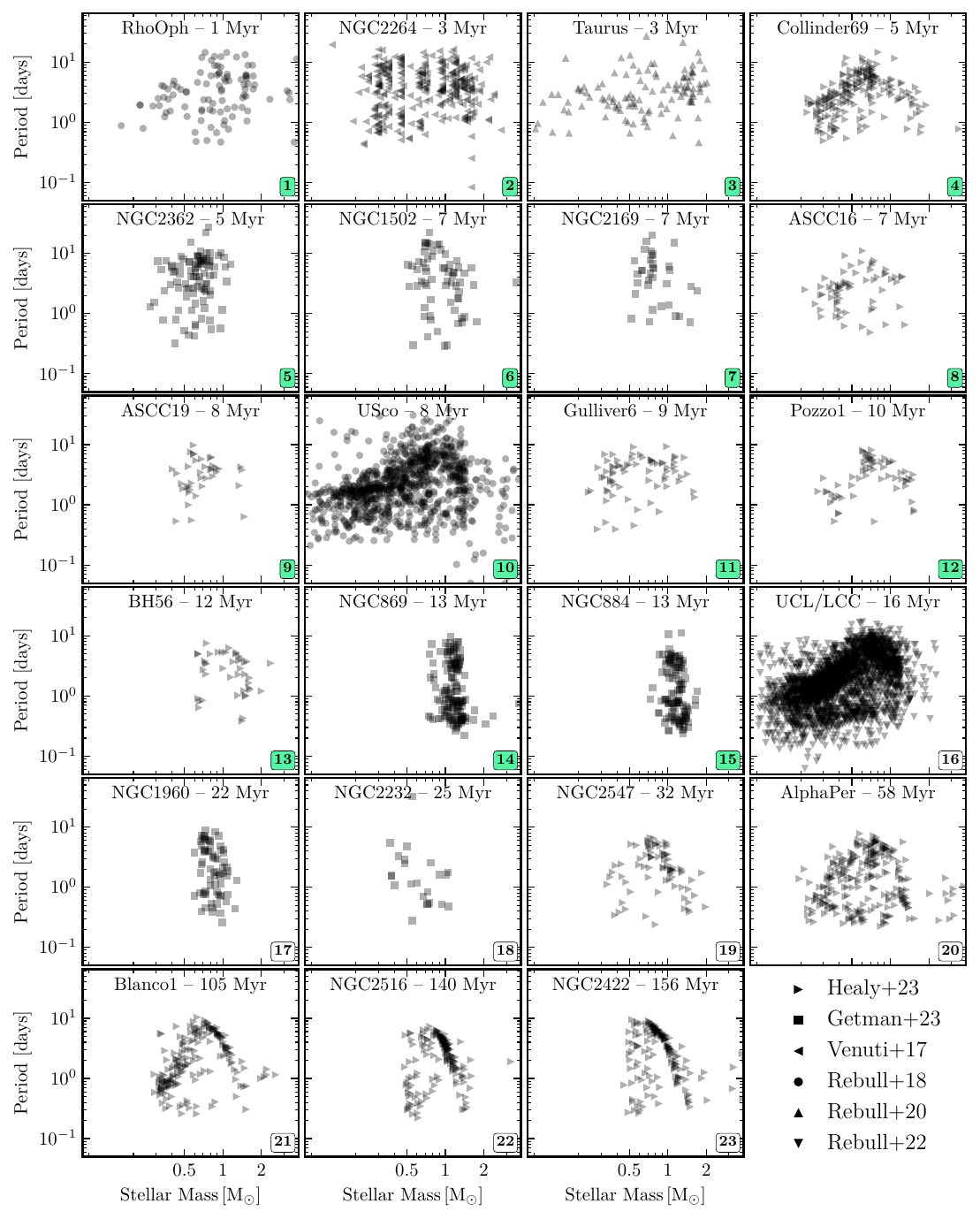}
        \caption{Stellar rotation period as a function of stellar mass for all analysed clusters listed in Tab.~\ref{tab:tab cl} (panel numbers correspond to cluster IDs). Stellar masses are derived via interpolation of PARSEC isochrones. Clusters included in the comparison with numerical simulations are marked with green labels. A broad dispersion in rotation periods is evident at young ages, which progressively narrows in older clusters (e.g.~NGC\,2422), indicating that the early diversity in stellar spins converges with time. Most clusters also show a characteristic break at the high-mass end, beyond which rotation periods decrease; this feature is particularly prominent in Blanco\,1.}
        \label{fig:all clusters}
    \end{figure*}

    Here we present the stellar rotation vs.~mass relation for the analysed young Galactic star clusters. Including data from all the sources discussed in the previous section, the sample of clusters consist of 23~clusters with ages between $1\,\myr$ to $500\,\myr$. Rotation periods vs.~masses are displayed in Fig.~\ref{fig:all clusters}, while the adopted cluster properties, e.g.~age, distance and source, are listed in Table~\ref{tab:tab cl}.

    As described in Sec.~\ref{sec:sim}, the stars in our simulations have maximum ages of $1\,\myr$, with most of the stars being younger than $0.5\,\myr$. Therefore, to ensure a proper comparison with observed clusters, we further restricted the sample of clusters to those younger than $15\,\myr$, highlighted in green in Fig.~\ref{fig:all clusters}. The final sample of clusters hence includes 15~clusters.
        
    Stars in the analysed clusters have masses between 0.3 and approximately 2\,$\msun$, and rotational periods ranging from about 0.1 to more than 10~days. The dataset from \citet{getman2023} covers a narrower mass range, including only stars more massive than about $0.5\,\msun$, with the exception of NGC\,2362.
    
    Additionally, as clusters become older, the relationship between rotational period and stellar mass tightens. This is clearly observed in NGC\,2422, the oldest cluster in our sample ($\sim 160\,\myr$), which displays a well-defined relationship. In contrast, younger clusters show a larger spread in rotational periods, though Collinder\,69 may represent an exception.

    Finally, we note that the overall range of the period–mass distributions of the clusters included in this study are remarkably similar to one another, showing only weak evolution. This uniformity likely reflects that most of the rotational evolution occurs very early, within the first $\lesssim1,\myr$. At later stages ($\gtrsim10$–$100,\myr$), the evolution becomes modest, leading to only small differences in the average rotation periods and their spreads among the young clusters analysed here. The next significant phase of evolution becomes apparent only at much older ages, e.g.~$>300\,\myr$, as shown in Fig.~1 of \citet{bouma2023}.

    \section{Comparison between simulations and observations} \label{sec:discussion}
    
    The aim of this study is to investigate how simulations with different physical conditions (described in Sec.~\ref{sec:sim}) compare to observations of stellar rotation. We specifically compare stellar spin distributions from MHD simulations of star-cluster formation \citep[see][for a detailed description of the simulations; see simulation parameters in Tab.~\ref{tab:tab sim}]{mathew2025} with observed rotational periods in young star clusters (Tab.~\ref{tab:tab cl}), focusing on the relationship between stellar mass and rotation period.
    
    As mentioned in Sec.~\ref{sec:sim}, the simulations cover timescales shorter than $1\,\myr$ and thus represent the earliest stages of star and cluster formation. Thus, it is important to emphasise that the simulated stars are approximately one order of magnitude younger than any of the observed clusters.

    \begin{figure*}
        \centering
        \includegraphics[width=0.99\textwidth]{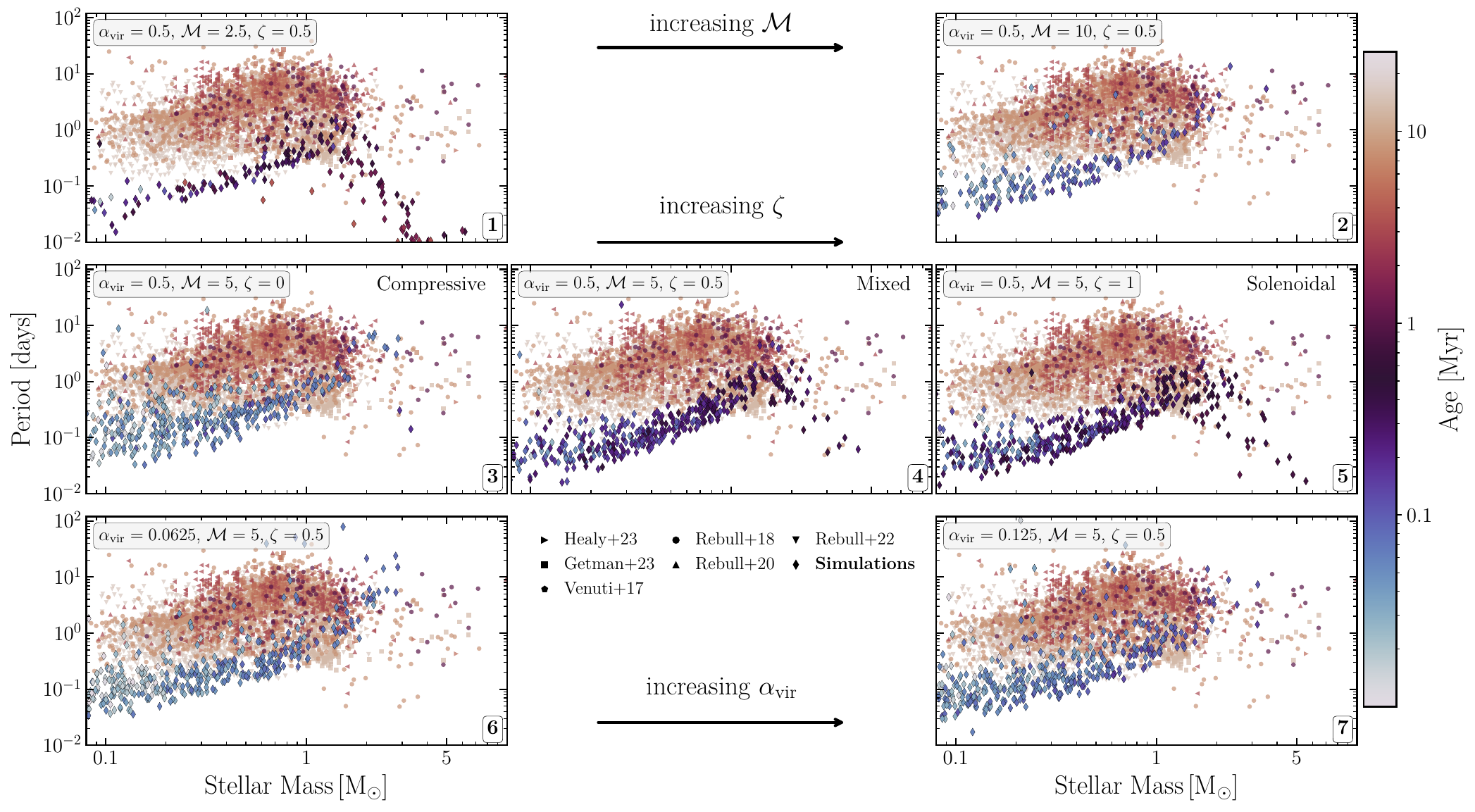}
        \caption{Updated. Stellar rotation period vs.~stellar mass for the simulations shown in Fig.~\ref{fig:sim} (Tab.~\ref{tab:tab sim}) together with the observations shown in Fig.~\ref{fig:all clusters} (Tab.~\ref{tab:tab cl}). Observations and simulations are displayed with different markers, and colour-coded by stellar/cluster age. Simulations reproduce the increasing period–mass trend at low masses but are typically faster than observations by roughly an order of magnitude, likely due to the age mismatch between very young simulated stars ($\lesssim1\,\myr$) and older observed populations ($1-15\,\myr$). Certain setups (e.g.~compressive driving in panel, lower $\avir$) exhibit a better agreement with the data in terms of period-mass values, spread in periods, with the high-mass break in simulations occurring at higher masses than observed.}
        \label{fig:sim comparison}
    \end{figure*}

    Figure~\ref{fig:sim comparison} compares our simulations with observed clusters younger than $15\,\myr$. The simulations are the same as in Fig.~\ref{fig:sim}, except that we only show stars with masses above $0.1\,\msun$, since the observations do not include lower-mass stars. In the figure, the top row displays variations in the mass-period relation for varying Mach numbers ($\mach$), increasing from left to right. The middle row illustrates the same but for different turbulence driving modes ($\zeta$), ranging from fully compressive ($\zeta=0$, left) to fully solenoidal ($\zeta=1$, right). The bottom row shows simulations with varying virial parameter ($\avir$). Observed clusters (see legend) and simulations (diamonds) are coloured according to stellar age, with simulated data appearing in darker blue or black colours.
    
    Since our simulations represent a younger evolutionary phase compared to the observed clusters, our primary focus is on the shape of the mass–period relationship rather than on the absolute values of rotational periods. Nevertheless, it is worth noting that simulations with compressively-driven turbulence (panel~3) and $\avir=0.125$ (panel~7) show rotational periods closest to the observed values, although even in this case, the simulated stars still rotate roughly one order of magnitude faster than the observed stars. Additionally, simulations with a virial parameter lower than the standard value of $\avir=0.5$ show a larger spread in rotational periods and slower rotation rates.
    
    All simulations qualitatively reproduce the observed general trend for low-mass stars, where periods increase (rotation slows) toward higher stellar masses. However, the behaviour of more massive stars varies significantly depending on the physical conditions adopted in the simulations. Simulations in panels~1, 4, and 5 display a notable break around 1--$2\,\msun$, beyond which periods decrease (rotation speeds up) as mass increases. The break and transition to a negative correlation in the mass–period relation is likely due to runaway accretion in high-mass stars, as discussed in Section~\ref{sec:sim}. Specifically, the break is seen only under specific parameter combinations: \textit{(i)} a low Mach number ($\mach=2.5$) together with a standard virial parameter ($\avir=0.5$) and mixed turbulence driving ($\zeta=0.5$; panel~1); \textit{(ii)} the base model with $\mach=5$, $\avir=0.5$, and $\zeta=0.5$ (panel~4); and \textit{(iii)} solenoidal turbulence driving ($\zeta=1$) combined with $\mach=5$ and $\avir=0.5$ (panel~5). Changing any of these parameters in isolation does not produce the break within our grid. A common characteristic of these simulations is that the clouds and the embedded star clusters are relatively older. Since these specific simulations also exhibit relatively low star formation rates, they had to be evolved significantly to reach a star formation efficiency of 5\%, which is when the stellar properties are measured for this study. This means that the clouds are more dynamically evolved and gravity had sufficient time to focus local gas towards the potential well, where the high-mass stars are generally located. As a result of the ample supply of gas around the stars, they continue accreting for extended periods, which can result in spin-up and a decrease in rotation period. Remarkably, a similar trend is also observed in several observed clusters (e.g.~Collinder\,69, NGC\,1502, ASCC\,16, USco, Pozzo\,1, BH\,56, and all older clusters), although in the observations the break occurs at lower stellar masses, approximately between 0.5 and $1\,\msun$.
        
    While the observations display a wide range of rotation periods for any given mass, the simulations show a narrow and well-defined trend, with the exception of low-mass stars in panel~3, which displays a larger scatter in rotational periods, especially for mass $\lesssim0.5\,\msun$.

    The fact that the rotation periods of stars in all of the simulations are approximately one order of magnitude shorter than in the observed clusters is primarily a consequence of their much younger ages. The simulated populations probe the very earliest, deeply embedded stages of star formation ($\lesssim1\,\myr$), when stars are expected to rotate more rapidly, before significant angular-momentum loss through disc–star coupling occurs. Additional factors may further contribute to the offset. First, there is a spatial–scale mismatch: angular momentum in the simulations is measured at the sink particle scale ($r_{\rm sink}=250\,\au$), while the observations probe stellar surface rotation. Second, although jet and outflow feedback with angular-momentum transport is included in the simulations \citep{federrath2014}, these models remain approximate and may not capture the full efficiency of angular-momentum removal. Finally, processes acting between the stellar surface and the inner disc, such as magnetic braking, magnetospheric ejections, and disc locking \citep[e.g.][]{amard2019}, are unresolved, likely contributing to the systematically shorter periods seen in simulations.

    Nonetheless, when comparing the simulated and observed samples, we find consistent physical trends: rotation periods increase with stellar mass, and a high-mass break appears in some of the simulation models. Additionally, the relative differences among simulation setups (e.g.~solenoidal versus compressive driving, or variations in $\avir$) provide direct insight into how cloud-scale initial conditions shape early stellar angular-momentum distributions.

    \subsection{Early stellar period evolution} \label{sec:earlyevolution}
    
    To test whether the age difference between simulations and observations can account for the offset in rotation periods, we examine how stellar rotation evolves with age within the simulations. Here, a star’s age is defined as the time elapsed between its formation and the end of the simulation.
    
\begin{figure*}
        \centering
        \includegraphics[width=0.99\textwidth]{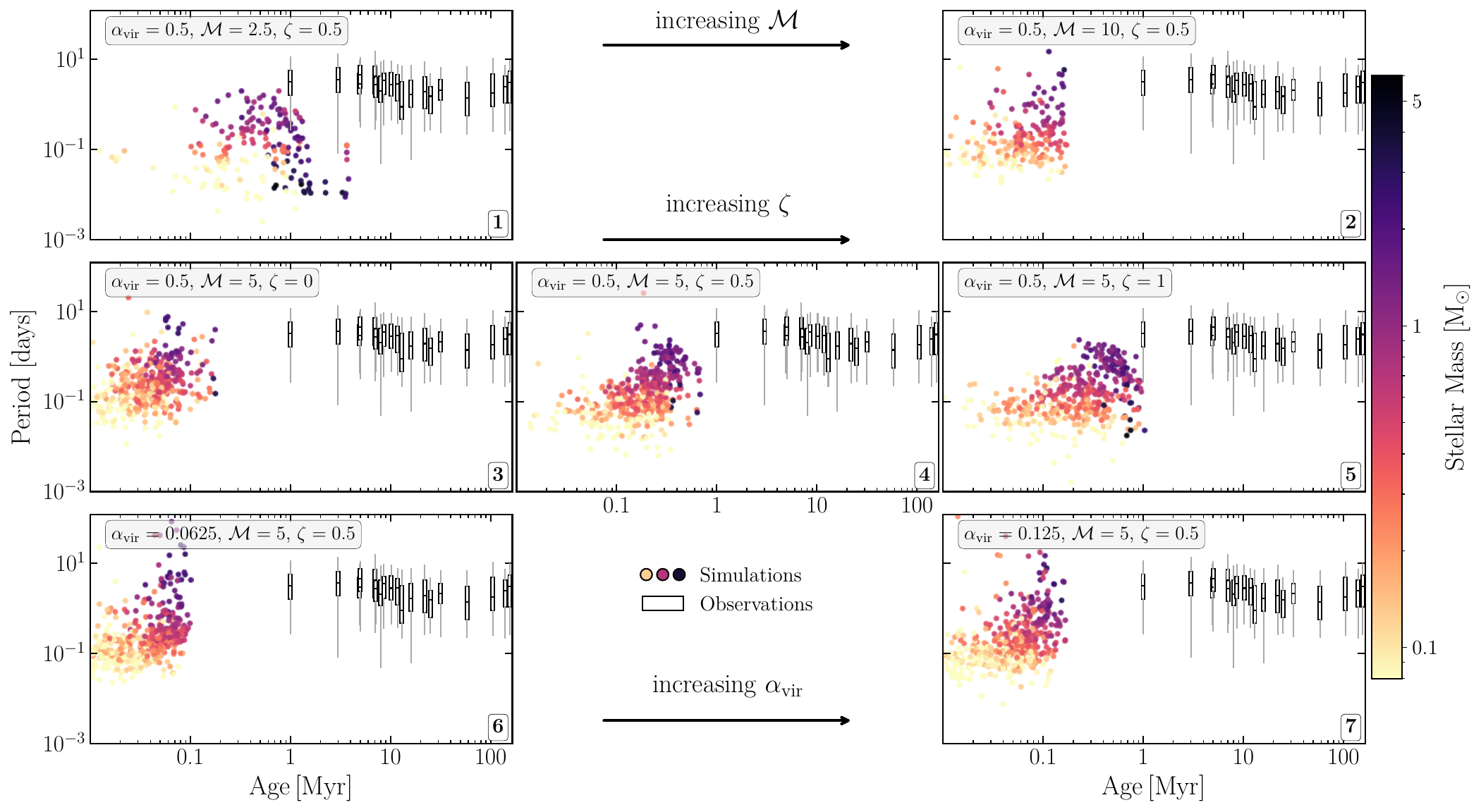}
        \caption{Stellar rotation period versus age for simulations and observations. Simulated stars are shown as circles, colour-coded by stellar mass. Observed clusters are represented by boxplots, where boxes and whiskers indicate the interquartile range (IQR) and 1.5 times the IQR, respectively, and horizontal lines mark the median period values. In all simulation setups, rotation periods increase with stellar age, indicating that stars spin down during the first $0.1$–$1\,\myr$. 
        On average, the oldest stars in each simulation rotate about an order of magnitude more slowly than the youngest ones, implying substantial angular-momentum loss at early times, consistent with the discussion in Sec.~\ref{sec:discussion}.}
        \label{fig:period_age}
    \end{figure*}

    Figure~\ref{fig:period_age} shows rotation period as a function of stellar age for all simulated stars, colour-coded by stellar mass, together with boxplots representing the observed clusters. In every simulation, older stars rotate more slowly, with periods increasing by nearly an order of magnitude within the first $0.1$–$1\,\myr$. A mild dependence on mass is also apparent: as stars accrete and grow in mass, they tend to rotate more slowly.
    
    These trends indicate that substantial angular momentum loss occurs very early, consistent with expectations from magnetic braking, outflows, and early star–disc coupling. Two caveats are worth noting. First, there is intrinsic scatter in the period at fixed age, reflecting variations in accretion histories and local environments. Second, even the oldest simulated stars still rotate faster than those in the youngest observed clusters. Taken together, these results suggest that (i) most angular-momentum evolution takes place within the first $\myr$, and (ii) the age difference is the main driver of the absolute offset between simulations and observations, with additional contributions from unresolved star–disc coupling in the models.

    \subsection{Quantitative comparison between simulations and observations: estimating angular momentum loss}\label{sec:momentum_loss}
    To quantify the total angular-momentum loss required to match the simulations with the observed clusters, we apply a constant rescaling factor $\fdl$ to the simulated angular momenta. Specifically, for every star we replace $L \rightarrow L\fdl$ (i.e. we uniformly reduce angular momenta by the factor $\fdl$, which correspondingly increases rotation periods by $1/\fdl$ given $P\propto I/L$). This rescaling is meant to include both the age-related period evolution between the simulations and the clusters, and additional loss from unresolved mechanisms in the simulations (e.g.~star–disc coupling, magnetospheric ejections, or wind braking). In principle, the required angular-momentum loss could depend on stellar mass (e.g. via mass-dependent disc lifetimes or magnetic coupling efficiencies), but modelling and fitting a mass-dependent $\fdl(M_\star)$ would require additional assumptions and is beyond the scope of this work; here we adopt a single, constant $\fdl$ for all stars as a first-order approximation.

\begin{figure*}
        \centering
        \includegraphics[width=0.99\textwidth]{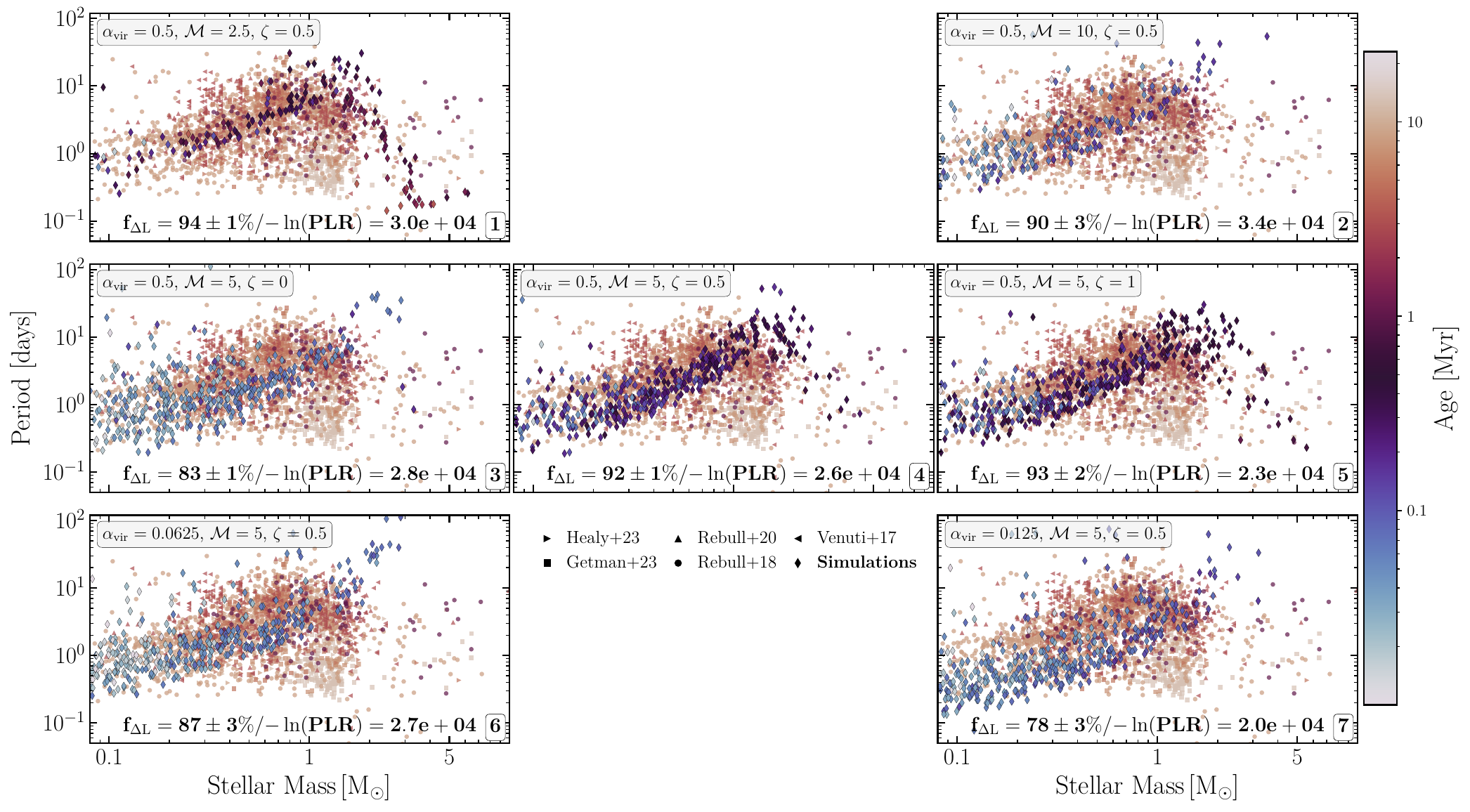}
        \caption{Same as Fig.~\ref{fig:sim comparison} but including only clusters younger than $15\,\myr$, after scaling simulated periods to best match the observational distributions. The inset reports the best-fit angular-momentum loss fraction $\fdl$ (and uncertainty) and the corresponding $-\lnplr$. The fits imply $\sim78$–$94\%$ angular-momentum must be lost between the $\sim0.1-1\,\myr$ of the simulations and the few-$\myr$-old observed clusters, with $\avir=0.125$ (panel 7) and solenoidal driving (panel 5) setups providing the best overall matches.}
        \label{fig:momentum loss}
    \end{figure*}

    Throughout this analysis, the observational sample is restricted to clusters younger than $15\,\myr$ (Table~\ref{tab:tab cl}). We explore $\fdl\in[60\%,\,99\%]$\footnote{The lower limit is chosen because smaller values fail to produce acceptable agreement.}. For each rescaled simulation, we evaluate the agreement with the observed period–mass distributions using the Poisson likelihood ratio (PLR) statistic \citep[][]{cash1979}, defined as
    \begin{equation}
    -\lnplr = \sum_{i,j} \left[ m_{ij} - n_{ij} + n_{ij} \ln\!\left(\frac{n_{ij}}{m_{ij}}\right) \right],
    \label{eq:plr}
    \end{equation}
    where $n_{ij}$ and $m_{ij}$ are the observed and model counts in the $(i,j)^\mathrm{th}$ bin of the two-dimensional histogram in stellar mass and rotation period, respectively. We compute the PLR using two-dimensional histograms in stellar mass and rotation period with $10\times15$ logarithmically-spaced bins\footnote{The number of bins is chosen as a compromise between resolution and statistics; alternative binning yields consistent results.}. For each simulation, we minimise the PLR to obtain the best-fit value of $\fdl$, and use \texttt{emcee} \citep{emcee} to estimate posterior distributions and uncertainties. Best-fit values and their associated $-\lnplr$ are reported in Fig.~\ref{fig:momentum loss}.

    We find best-fit $\fdl$ values ranging from $\sim84\%$ to $\sim94\%$. The simulations with solenoidal turbulence driving ($\avir=0.5;~\mach=5;~\zeta=1$, panel~5) and with low virial parameter ($\avir=0.125;~\mach=5;~\zeta=0.5$, panel~7) provide the best overall agreement, with very similar PLR values. Both reproduce the observed trends and spreads in the period–-mass distributions. These two simulations remain the best performers even when alternative models for the disc mass correction are applied (see Sec.~\ref{sec:sim} and App.~\ref{appendix:disc}; Tab.~\ref{tab:disc}). 
    
    Quantitatively, a good match to the observations is achieved when the stellar angular momenta in the solenoidal driving case are scaled down by $\sim93\%$, and when those in the $\avir=0.125$ case are scaled down by $\mathbf{\sim78\%}$. It is also worth mentioning that the simulation with $\avir=0.125$ consistently requires a lower rescaling factor with respect to the other simulation setups.
    As an additional validation step, we compared simulations and observations using complementary statistical metrics, including the Kolmogorov–Smirnov test \citep{kolmogorov1933,smirnov1948}, the Anderson–Darling test \citep{anderson1952}, and the two-dimensional energy distance metric \citep{szekely2016energy}, implemented via \texttt{SciPy} \citep{scipy}. These tests yield results qualitatively consistent with the PLR analysis, supporting our conclusions.

    Finally, we note that the inferred $\fdl$ values are qualitatively consistent with the scale-dependent relation $j \propto r^{0.3}$ reported by \citet{gaudel2020}. Their analysis shows only a weak dependence of angular momentum on scales between $50$ and $1600\,\au$, which indicates that adopting a sink radius of $250\,\au$ does not introduce a significant bias in our simulations.

    \section{Summary and conclusions}\label{sec:conclusion}
    
    In this work, we investigated stellar spin distributions in very young star clusters by comparing rotation periods obtained from state-of-the-art MHD simulations of star-cluster formation with observational data from clusters younger than $15\,\myr$. The simulations incorporate key physical processes, including turbulence-driven fragmentation, magnetic fields, stellar radiative heating, outflows, and gravitational collapse, and thus provide predictions for the earliest phases of stellar spin evolution.
    
    Our comparison shows that the simulations recover the overall trend of increasing rotation period with stellar mass at the low-mass end and, in several setups, a break in the mass–period relation at higher masses. In the simulations, this break occurs at somewhat higher masses (and can be absent in the very youngest model clusters), which we interpret as a signature of continued accretion and spin-up of massive stars in dynamically young environments. This behaviour suggests that the emergence and location of the break are time-dependent diagnostics of angular-momentum evolution during the first $\myr$ (Fig.~\ref{fig:sim}).

    The stars in our simulations are significantly younger ($\lesssim1\,\myr$) than those in the observational comparison sample (typically $1$–$15\,\myr$). This age difference, together with the associated evolution of star–disc systems, is the main factor behind the systematic offset, whereby the simulated young stars rotate faster than the observed (older) stars, often by roughly an order of magnitude (Fig.~\ref{fig:sim comparison}). Specifically, the simulations show that stars already undergo significant spin-down early on, losing a substantial fraction of their angular momentum within the first  $0.1-1\,\myr$. This behaviour supports the interpretation that most angular-momentum evolution occurs very early in the stellar lifetime, well before the ages sampled by current observational data.
    When we quantify this discrepancy by rescaling the simulated angular momenta to match clusters younger than $15\,\myr$, we find that approximately $80-95\%$ of the initial angular momentum must be removed within the first $\sim0.1-1\,\myr$ (Fig.~\ref{fig:momentum loss}).
    
    While unresolved physics in the simulations (e.g.~magnetospheric disc locking and magnetic or wind braking operating between the stellar surface and the inner disc) likely contributes to the remaining differences, our results indicate that evolutionary timing is the dominant factor. Nevertheless, a realistic treatment of angular-momentum transport remains crucial: without it, simulations cannot capture the full range of mechanisms that regulate stellar spin. Incorporating physically motivated sub-grid prescriptions for these processes—and extending simulations to later evolutionary stages—will be essential for reproducing both the absolute period scales and the time-evolving shape (including the break) of the observed mass–period relation in young clusters. Taken together, the results point to substantial angular-momentum loss occurring very early, from collapse through the Class~II stage ($\sim1\,\myr$), consistent with classical expectations (e.g.~\citealt{bodenheimer1995}).
    
    \section*{Acknowledgements}
    We thank the anonymous referee for their useful comments which significantly improved the quality of the manuscript. C.~F.~acknowledges funding provided by the Australian Research Council (Discovery Projects DP230102280 and DP250101526), and the Australia-Germany Joint Research Cooperation Scheme (UA-DAAD). We further acknowledge high-performance computing resources provided by the Leibniz Rechenzentrum and the Gauss Centre for Supercomputing (grants~pr32lo, pr48pi and GCS Large-scale project~10391), the Australian National Computational Infrastructure (grant~ek9) and the Pawsey Supercomputing Centre (project~pawsey0810) in the framework of the National Computational Merit Allocation Scheme and the ANU Merit Allocation Scheme.
    
    \section*{Data Availability}
    The data underlying this article are available on reasonable request from the corresponding author. The observational datasets used are publicly available from \citet{venuti2017, healy2023, getman2023, rebull2018, rebull2020, rebull2022}.

\bibliographystyle{mnras}
\bibliography{main,federrath}
    
\appendix

    \section{disc mass correction model and sensitivity tests} \label{appendix:disc}

    To determine the stellar mass component from unresolved star--disc systems in our simulations, we applied a post-processing correction based on a toy model for the disc-to-star mass ratio, $\mu(M_\star, t)=M_\mathrm{disc}/M_\star$. This model accounts for the observed trend that disc lifetimes depend on stellar mass---discs around high-mass stars dissipate more rapidly, while those around low-mass stars tend to persist longer.

    We define the mass ratio $\mu$ as
    \begin{equation}
    \mu(M_\star, t) = \mu_0 \exp\left(-\frac{t}{\tau_{\rm eff}(M_\star)}\right),
    \end{equation}
    where $t$ is the stellar age in $\myr$, $\mu_0=0.8$ is the initial disc-to-star mass ratio, and $\tau_{\rm eff}$ is a mass-dependent effective disc lifetime. In our model, $\tau_{\rm eff}$ increases linearly with stellar mass, ranging from $\tau_\mathrm{min}=1.0\,\myr$ at $0.1\,\msun$ to $\tau_\mathrm{max}=2.0\,\myr$ at $1.5\,\msun$. For stars less massive than $0.1\,\msun$ and more massive than $1.5\,\msun$, the disc timescale is fixed to $\tau_\mathrm{min}$ and $\tau_\mathrm{max}$, respectively. The exponential decay form of $\mu(t)$ captures the observed decline of infrared excess and accretion signatures with age across a wide range of stellar masses \citep{mamajek2009, fedele2010}. The adoption of a mass-dependent timescale is further motivated by evidence that disc survival correlates with stellar mass \citep[e.g.][]{ribas2015}.
    
    We then determined the stellar mass $M_\star$ from the total system mass $M_\mathrm{sys}$ by iteratively solving the equation
    \begin{equation}
    M_\mathrm{sys} = M_\star \left[1 + \mu(M_\star, t)\right].
    \end{equation}
    
\begin{figure}
        \centering
        \includegraphics[width=0.48\textwidth]{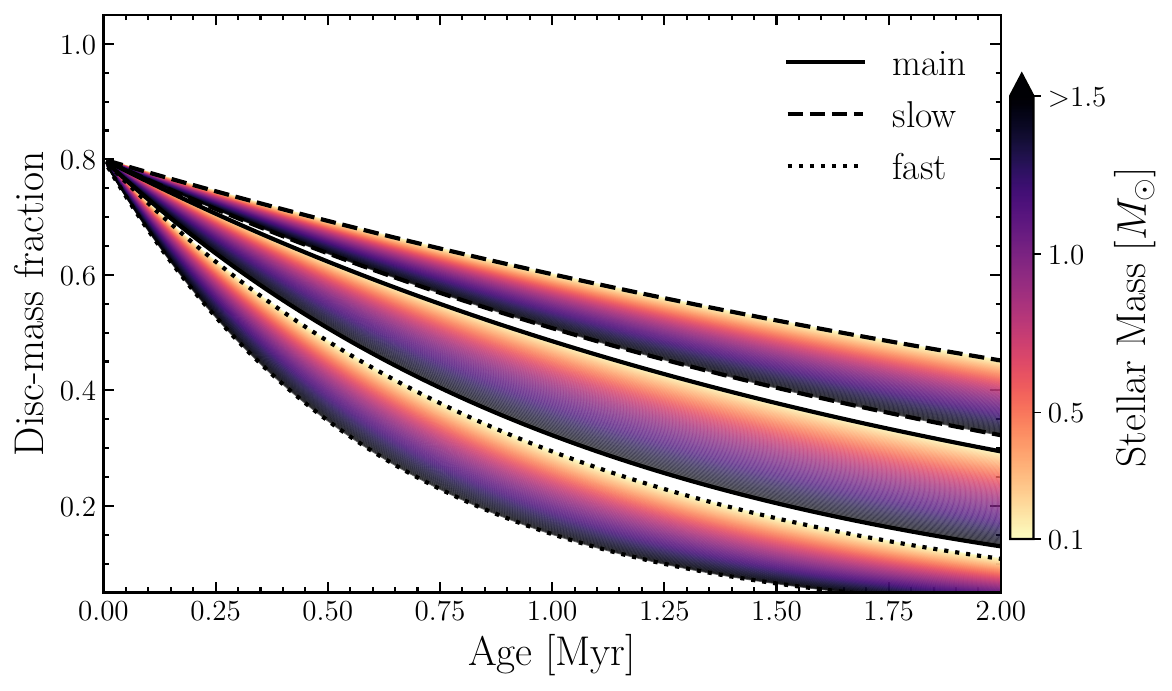}
        \caption{Fraction of disc mass vs.~stellar age for different stellar masses across the tested disc-fraction models. The main model used in the analysis is shown as a solid black line; coloured curves indicate different stellar masses. The inferred angular-momentum rescaling factors (Sec.~\ref{sec:momentum_loss}, Tab.~\ref{tab:disc}) are only weakly sensitive to the disc mass prescription, indicating that our conclusions are robust to plausible variations in disc evolution.}
        \label{fig:mass correction}
    \end{figure}

    To test the robustness of our results, we varied the minimum and maximum disc lifetimes to model slower ($\tau_\mathrm{min}/\tau_\mathrm{max} = 2.1/3.5$) and faster ($\tau_\mathrm{min}/\tau_\mathrm{max} = 0.5/1.0$) disc dissipation. We also considered an extreme case where all stars were assigned a fixed disc fraction of 50\%, corresponding to a uniform $\mu = 0.5$ regardless of stellar mass or age. The different models are shown in Fig.~\ref{fig:mass correction} with distinct line styles. Our results, as well as the discussion in Sec.~\ref{sec:momentum_loss}, are not significantly affected by the choice of mass correction. The best-fit fractions of angular momentum loss, derived as described in Sec.~\ref{sec:momentum_loss}, are reported in Tab.~\ref{tab:disc}.
        
\begin{table}
    \centering
    \begin{tabular}{c|ccccc}
        \hline
        Disc model & $\avir$ & $\mach$ & $\zeta$ & $\fdl$ [\%] & $-\lnplr$ \\
        \hline
        \textbf{Main} &  &  &  &  &  \\
        & 0.5    &  2.5  & 0.5 & $94 \pm 1$ & $3.0\times10^4$ \\
        & 0.5    & 10.0  & 0.5 & $90 \pm 3$ & $3.4\times10^4$ \\
        & 0.5    &  5.0  & 0.0 & $83 \pm 1$ & $2.8\times10^4$ \\
        & 0.5    &  5.0  & 0.5 & $92 \pm 1$ & $2.6\times10^4$ \\
        & 0.5    &  5.0  & 1.0 & $93 \pm 2$ & $2.3\times10^4$ \\
        & 0.0625 &  5.0  & 0.5 & $87 \pm 3$ & $2.7\times10^4$ \\
        & 0.125  &  5.0  & 0.5 & $78 \pm 4$ & $2.0\times10^4$ \\
        \textbf{Slow} &  &  &  &  &  \\
        & 0.5    &  2.5  & 0.5 & $92 \pm 1$ & $2.6\times 10^4$ \\
        & 0.5    & 10.0  & 0.5 & $90 \pm 2$ & $3.4\times 10^4$ \\
        & 0.5    &  5.0  & 0.0 & $85 \pm 1$ & $2.7\times 10^4$ \\
        & 0.5    &  5.0  & 0.5 & $91 \pm 1$ & $2.5\times 10^4$ \\
        & 0.5    &  5.0  & 1.0 & $93 \pm 2$ & $2.2\times 10^4$ \\
        & 0.0625 &  5.0  & 0.5 & $87 \pm 3$ & $2.6\times 10^4$ \\
        & 0.125  &  5.0  & 0.5 & $78 \pm 3$ & $2.1\times 10^4$ \\
        \textbf{Fast} &  &  &  &  &  \\
        & 0.5    &  2.5  & 0.5 & $93 \pm 2$ & $3.3\times 10^4$ \\
        & 0.5    & 10.0  & 0.5 & $89 \pm 1$ & $3.5\times 10^4$ \\
        & 0.5    &  5.0  & 0.0 & $88 \pm 1$ & $2.6\times 10^4$ \\
        & 0.5    &  5.0  & 0.5 & $91 \pm 1$ & $2.8\times 10^4$ \\
        & 0.5    &  5.0  & 1.0 & $92 \pm 2$ & $2.4\times 10^4$ \\
        & 0.0625 &  5.0  & 0.5 & $85 \pm 1$ & $2.5\times 10^4$ \\
        & 0.125  &  5.0  & 0.5 & $77 \pm 3$ & $2.0\times 10^4$ \\
        \hline
    \end{tabular}
    \caption{Best-fit angular momentum loss resulting from the comparison between simulations and observations, as discussed in Sec.~\ref{sec:momentum_loss}, for the three disc mass fraction models: Main, Fast, and Slow. The column denoted with $\fdl$ indicates the best-fit angular momentum loss and its uncertainty (in percent), while the last column is the negative logarithm of the Poisson likelihood ratio corresponding to the best-fit $\fdl$.}
    \label{tab:disc}
\end{table}

\end{document}